\documentclass[aps,prb,twocolumn,superscriptaddress,showpacs,floatfix]{revtex4}

\usepackage{graphicx}
\graphicspath{{figs/}}
\begin{document}
\title{Young's modulus of Graphene: a molecular dynamics study}
\author{Jin-Wu~Jiang}
    \affiliation{Department of Physics and Centre for Computational Science and Engineering,
             National University of Singapore, Singapore 117542, Republic of Singapore }
\author{Jian-Sheng~Wang}
    \affiliation{Department of Physics and Centre for Computational Science and Engineering,
                 National University of Singapore, Singapore 117542, Republic of Singapore }
\author{Baowen~Li}
        \affiliation{Department of Physics and Centre for Computational Science and Engineering,
                     National University of Singapore, Singapore 117542, Republic of Singapore }
        \affiliation{NUS Graduate School for Integrative Sciences and Engineering,
                     Singapore 117456, Republic of Singapore}

\date{\today}
\begin{abstract}
The Young's modulus of graphene is investigated through the
intrinsic thermal vibration in graphene which is `observed' by
molecular dynamics, and the results agree very well with the recent
experiment [Science \textbf{321}, 385 (2008)]. This method is
further applied to show that the Young's modulus of graphene: (1).
increases with increasing size and saturates after a threshold value
of the size; (2). increases from 0.95 TPa to 1.1 TPa as temperature
increases in the region [100, 500]K; (3). is insensitive to the
isotopic disorder in the low disorder region ($< 5\%$), and
decreases gradually after further increasing the disorder
percentage.
\end{abstract}

 \maketitle

The single layer graphene has unique electronic and other physical
properties, thus becoming a promising candidate for various device
applications.\cite{Novoselov1, Neto} Among others, excellent
mechanical property is an important advantage for the practical
applications of graphene. Experimentally, the Young's modulus ($Y$)
of graphene has been measured by using atomic force microscope (AFM)
to introduce external strain on graphene and record the
force-displacement relation.\cite{Lee} The measured value for
Young's modulus is $1.0\pm 0.1$ TPa in this experiment.
Theoretically, the Young's modulus of graphene can be studied in a
parallel way. Once the external strain is applied on graphene, the
internal force or potential can be calculated in different
approaches, such as \textit{ab initio} calculations,\cite{Kudin,
Lier, Konstantinova} molecular dynamics (MD)\cite{Khare} and
inter-atomic potentials.\cite{Reddy, Huang, Lu} Then the Young's
modulus can be obtained from the force-displacement or the
potential-displacement relation. For the carbon nanotubes (CNT), the
Young's modulus is theoretically studied in a similar way as that in
graphene. However, in the experiment, besides the AFM
method,\cite{Tombler} another group measured the Young's modulus of
CNT by  observing the thermal vibration at the tip of the CNT using
the transmission electron microscopy (TEM).\cite{Treacy, Krishnan}
For some unknown reasons, possibly technical challenges, this
experimental method does not appear in the study of the Young's
modulus in graphene. As a supplement to this vacancy, the present
work `observes' the thermal vibration of graphene by MD instead of
TEM, and then calculates the Young's modulus from the `observed'
thermal vibration.

In the engineering application of graphene, it will be beneficial if
the mechanical property of graphene can be adjusted according to the
demand. There are some possible methods that can manipulate the
value of Young's modulus in graphene, such as size of the sample,
temperature, isotopic disorder, etc. It is a matter of practical
importance and theoretical interest to find an effective method to
control the mechanical property of graphene. The present calculation
method for the Young's modulus of graphene in this paper is readily
applicable to address these issues.

In this paper, we investigate the Young's modulus of graphene by
`observing' the thermal vibrations with MD. The calculated Young's
modulus is in good agreement with the recent experimental one. Using
this method, we can systematically study different effects on the
Young's modulus: size, temperature and isotopic disorder. It shows
that the Young's modulus increases as graphene size increases, and
saturates. In the temperature range $100-500$ K, $Y$ increases from
0.95 TPa to 1.1 TPa as $T$ increases. For the isotopic disorder
effect, $Y$ keeps almost unchanged within low disorder percentage
($< 5\%$), and decreases gradually after further increasing disorder
percentage.

In graphene there are both optical
and acoustic vibration modes
in the $z$ direction. For the optical phonon modes, the frequency is about
850 cm$^{-1}$, which is too high to
be considerably excited under 500 K.
While the acoustic phonon mode is
a flexure mode with parabolic dispersion $\omega=\beta k^{2}$, which will be
fully excited even at very low temperature. So
 the thermal mean-square vibration amplitude (TMSVA) of graphene in the $z$
 direction is mainly attributed to the flexure mode under 500 K.
In this sense, we consider the contribution of the flexure mode to
TMSVA for an elastic plate in the following. The $x$ and $y$ axes lie in the plate,
and $z$ direction is perpendicular to the plate. For convenience and without losing
generality, we consider a square plate with length $L$.

The equation for oscillations in $z$ direction of a plate is\cite{Landau}:
\begin{eqnarray}
&&\rho\frac{\partial^{2}z}{\partial t^{2}}+\frac{D}{h}\Delta^{2}z  =  0,
\label{eq_fm}
\end{eqnarray}
where $D=\frac{1}{12}Yh^{3}/(1-\mu^{2})$. $\Delta$ is the two-dimensional Laplacian and $\rho$ is the density of the plate.
$Y$ and $\mu$ are the Young's modulus and the Poisson ratio, respectively.
$h$ is the thickness of the plate.
We apply fixed boundary condition in $x$ direction, and periodic
boundary condition in $y$ direction:
\begin{eqnarray}
z(t,x=0,y) & = & 0,\nonumber\\
z(t,x=L,y) & = & 0,\\
z(t,x,y+L) & = & z(t,x,y).\nonumber
\label{eq_boundary}
\end{eqnarray}
The solution for the above partial differential equation under these boundary conditions can be found in Ref.~\onlinecite{Polyanin}:
\begin{eqnarray}
\omega_{n} & = & k_{n}^{2}\sqrt{\frac{Yh^{2}}{12\rho(1-\mu^{2})}},\nonumber\\
z_{n}(t,x,y) & = & u_{n}\sin (k_{1}x)\cdot\cos (k_{2}y)\cdot\cos(\omega_{n}t),\\
\vec{k} & = & k_{1}\vec{e}_{x}+k_{2}\vec{e}_{y},\nonumber
\label{eq_eigen}
\end{eqnarray}
where $k_{1}=\pi n_{1}/L$ and $k_{2}=2\pi n_{2}/L$.

Using these eigen solution, the TMSVA for $n$-th phonon mode in $(x,y)$ at temperature $T$ can be obtained\cite{Krishnan}:
\begin{eqnarray}
\sigma_{n}^{2}(x,y)& = & 4k_{B}T\times\frac{12(1-\mu^{2})}{Yh^{2}V}\times\frac{1}{k_{n}^{4}}(\sin (k_{1}x)\cos (k_{2}y))^{2}.
\end{eqnarray}
We mention that for those modes with $k_{1}\not=0$ and $k_{2}=0$, we have a similar result $\sigma_{n}^{2}(x,y)=2k_{B}T\times\frac{12(1-\sigma^{2})}{Eh^{2}V}\times\frac{1}{k_{n}^{4}}(\sin k_{1}x)^{2}.$

The spatial average of the TMSVA over $x$ and $y$ is:
\begin{eqnarray}
\langle\sigma_{n}^{2}\rangle & = & \frac{1}{S}\int\int_{D}\sigma_{n}^{2}(x,y)dxdy\nonumber\\
\nonumber\\ & = & k_{B}T\times\frac{12(1-\mu^{2})}{Yh^{2}V}\times\frac{1}{k_{n}^{4}},
\end{eqnarray}
 where $k_{1}\not=0$ and $k_{2}\not=0$. $D$ is the field in
$x\in[0,L]$ and $y\in[0,L]$, and $S=L^{2}$ is the area of $D$. If
$k_{1}\not=0$ and $k_{2}=0$, $\langle\sigma_{n}^{2}\rangle$ turns
out to have the same expression as this general one.

Because all modes are independent at the thermal equilibrium state at temperature $T$, they contribute to the TMSVA incoherently. As a result, the TMSVA at temperature $T$ is given by:
\begin{eqnarray}
\langle\sigma^{2}\rangle & = & \sum_{n=0}^{\infty}\langle\sigma_{n}^{2}\rangle\nonumber\\
\nonumber\\ & = & k_{B}T\times\frac{12(1-\mu^{2})}{Yh^{2}V}\times\sum_{n=0}^{\infty}\frac{1}{k_{n}^{4}}.\nonumber\\
& = & k_{B}T\times\frac{12(1-\mu^{2})}{Yh^{2}V}\times\frac{2S^{2}}{\pi^{4}}\times C\nonumber\\
\nonumber\\ & = & 0.31\times\frac{(1-\mu^{2})S}{h^{3}}\times\frac{k_{B}T}{Y}.
\label{eq_sigma2}
\end{eqnarray}
The constant $C=\sum_{n_{1}=1}^{+\infty}\sum_{n_{2}=0}^{+\infty}\epsilon_{n_{2}}\frac{1}{\left(n_{1}^{2}+4n_{2}^{2}\right)^{2}}\approx 1.2507$, the major part of which is due to the first nonzero phonon mode with $(n_{1}, n_{2})=(1, 0)$.
$\epsilon_{n_{2}}=1$ for $n_{2}=0$, and $\epsilon_{n_{2}}=2$ for other $n_{2}=1,2,3,...$.

As a result, the Young's modulus of the graphene is:
\begin{eqnarray}
Y & = & 0.3\times\frac{S}{h^{3}}\times\frac{k_{B}T}{\langle\sigma^{2}\rangle}.
\label{eq_Y}
\end{eqnarray}
The Poisson ratio in graphene\cite{Blakslee, Daniel} $\mu=0.17$ has been used in this expression for the Young's modulus.
There is arbitrariness in the definition of
thickness $h$ of the one atom thick graphene sheet.
For convenience of comparison between our theoretical results and the experimental
ones, we choose $h$ to be 3.35~{\AA}, the inter-layer space in graphite, which is also
used in the experimental work.\cite{Lee}

\begin{figure}[htpb]
  \begin{center}
    \scalebox{1.0}[1.0]{\includegraphics[width=7cm]{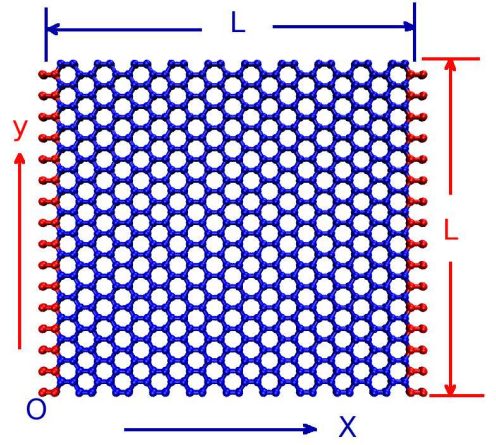}}
  \end{center}
  \caption{(Color online) Configuration of the graphene sample. The origin $O$ is at the left bottom of the sample. Two columns (red online) on the left and right sides are fixed. The length of the sample in this figure is $L=40$~{\AA}.}
  \label{fig_cfg}
\end{figure}

Fig.~\ref{fig_cfg} is the configuration of the graphene sheet in our simulation. The outmost two columns (red online) on the left and right sides are fixed during the simulation, while periodic boundary condition is imposed in the vertical direction. The origin of the coordinate is set at the left bottom of the sample. The $x$-axis is in the horizontal direction and $y$-axis is in the vertical direction.

The MD simulations are performed using the second-generation Brenner inter-atomic potential.\cite{Brenner}
The Newton equations of motion are integrated within the fourth order Runge-Kutta algorithm, in which a time step of 0.5 fs is applied. The typical MD simulation steps in this paper is $5\times 10^{5}$, corresponding to 0.25 ns simulation time.

The initial velocities of carbon atoms at temperature $T$ are assigned as independent Gaussian random variables drawn from the Maxwell-Boltzmann distribution. All atoms are at the optimized position at $t=0$. A long enough simulation time is used for the system to reach steady state. In our simulation, $5\times 10^{5}$ MD steps are used to ensure that the system has achieved the thermal equilibrium. Another $5\times 10^{5}$ MD steps are applied to calculate the time averaged quantities in this paper. The typical variation in the total energy of the system is very small ($< 2\%$).

After we obtain the $\langle\sigma^{2}\rangle$ from MD simulation, we can calculate the value of Young's modulus through Eq.~(\ref{eq_Y}). We note that the elastic theory has been successfully applied to describe atomic graphene system with about 400 carbon atoms.\cite{Cadelano} In this paper, the graphene samples have about 200--500 carbon atoms. So we expect the Eq.~(\ref{eq_Y}) resulted from elastic theory can also be applicable. To depress the possible error created by randomness in the simulation, we repeat 100 independent processes for each value of the Young's modulus in this work.
\begin{figure}[htpb]
  \begin{center}
    \scalebox{1.15}[1.15]{\includegraphics[width=7cm]{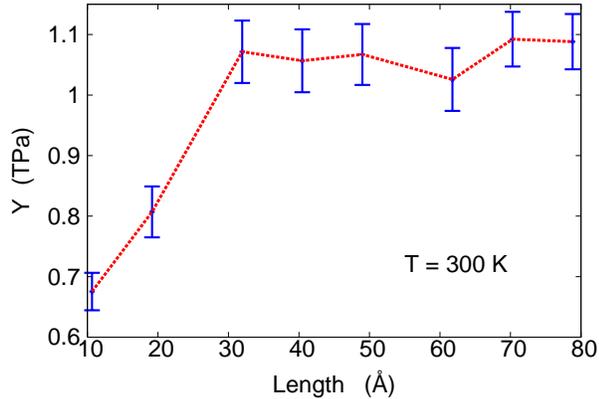}}
  \end{center}
  \caption{(Color online) The Young's modulus in graphene with different sizes.}
  \label{fig_size}
\end{figure}

Fig.~\ref{fig_size} shows the size dependence of the Young's modulus. When 10~{\AA} $<$ $L$ $<$ 40~{\AA}, $Y$ increases from 0.7 TPa to 1.1 TPa with increasing size, and this value (1.1 TPa) almost does not change with further increasing $L$. The increase of $Y$ with increasing size also shows up in some studies on the Young's modulus of CNT by various methods, where $Y$ increases with increasing diameter and reaches a saturate value.\cite{Robertson, ChangT, WangJB, Popov} In Fig.~\ref{fig_size}, the value of $Y$ in large size sample is 1.1 TPa. This value agrees quite well with the recent experimental $1 \pm 0.1$ TPa result.\cite{Lee}

\begin{figure}[htpb]
  \begin{center}
    \scalebox{1.15}[1.15]{\includegraphics[width=7cm]{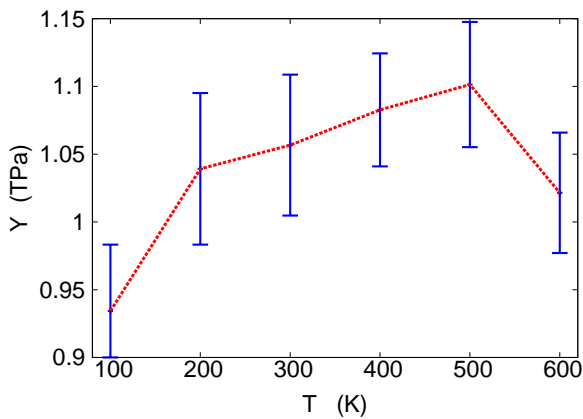}}
  \end{center}
  \caption{(Color online) The dependence of the Young's modulus on temperature $T$ for graphene with $L$=40~{\AA}.}
  \label{fig_tem}
\end{figure}
In Fig.~\ref{fig_tem}, we show the temperature dependence of the Young's modulus in the temperature range from 100 to 600 K. In the low temperature region $[100, 500]$K, $Y$ increases for 15$\%$ as $T$ increases. In the high temperature region $T>500$ K, $Y$ shows obvious decreasing behavior. This behavior indicates that the suitable temperature region for our method is $T<500$ K. If $T>500$ K, the optical phonon modes in the $z$ direction will also be excited together with the flexure mode, leading to a larger value for the TMSVA in our MD simulation. And the result from Eq.~(\ref{eq_Y}) will underestimate the value of Young's modulus.

\begin{figure}[htpb]
  \begin{center}
    \scalebox{1.15}[1.15]{\includegraphics[width=7cm]{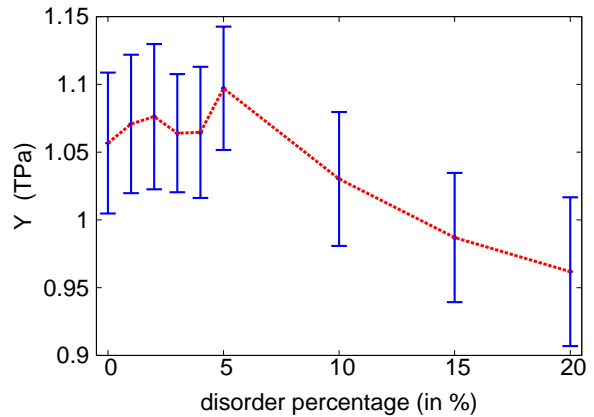}}
  \end{center}
  \caption{(Color online) The isotopic disorder effect on the Young's modulus of graphene at $T=300$ K with $L=40$~{\AA}.}
  \label{fig_isotope}
\end{figure}
Now we consider the result of the $^{14}$C isotopic disorder in the pure $^{12}$C graphene system. We expect this investigation of the isotopic disorder effect can give a useful clue to whether mechanical properties of graphene can be manipulated in this way. In our simulation, to calculate the value of $Y$ under a particular isotopic disorder percentage, $^{12}$C atoms are randomly substituted by certain number of $^{14}$C atoms. This procedure is done independently in each of our 100 simulation processes for one value of the Young's modulus. Results are shown in In Fig.~\ref{fig_isotope}. We find that the value of $Y$ remains almost unchanged for the low isotopic disorder percentage ($< 5\%$). Further increasing of the isotopic disorder percentage yields about 15$\%$ reduction of $Y$. This result tells us that the purification of the natural graphene can not obtain a higher value of $Y$. On the other hand, about 15 $\%$ reduction of $Y$ can be realized by increasing the isotopic disorder percentage. However, as 20$\%$ isotopic disorder only achieves 15$\%$ reduction of $Y$, it is not an effective method to control the value of $Y$ by modifying isotopic disorder percentage. This situation is very different from that in the thermal transport. The thermal conductivity has been shown to be very sensitive to the isotopic disorder percentage in the low disorder region with more than 40$\%$ reduction of thermal conductivity by less than 5$\%$ isotopic disorder percentage; while for higher disorder percentage, the thermal conductivity keeps almost unchanged.\cite{Anthony, Zhang, Chang} So the thermal conductivity can be greatly enhanced by synthesizing isotopically pure nanotubes.\cite{Chang}

In conclusion, we have used MD to obtain the thermal vibration of
graphene and then calculated the Young's modulus from the thermal
mean-square vibration amplitude. The advantage of this approach is
that we don't have to introduce external strain on the system, and
it can be easily applied to study different effects on the Young's
modulus. The theoretical results agree very well with the
experimental ones. As an application of this method, we study the
Young's modulus of graphene with different size. The temperature and
isotopic disorder effects on the Young's modulus are also
investigated. It shows that the Young's modulus increases with
increasing size when the graphene sample is smaller than 40~{\AA},
and reaches a saturated value in samples larger than 40~{\AA}. The
value of $Y$ increases from 0.95 TPa to 1.1 TPa as $T$ increases
from 100 K to 500 K. For the isotopic disorder effect, $Y$ keeps
almost unchanged in the low disorder region ($< 5\%$), and decreases
gradually for $15\%$ after increasing the disorder percentage up to
$20\%$. This finding provides the information that the isotopic
disorder is not an effective method to control the Young's modulus
of graphene.

We should point out that why we use the constant value of Poisson
ratio $\mu=0.17$. Actually $\mu$ also depends on the size,
temperature and isotopic doping. By applying external strain
($\varepsilon_{x}$) on the graphene in $x$ direction, and using MD
to record the resulted strain in $y$ direction ($\varepsilon_{y}$)
under different environment, i.e. different size, temperatures, or
isotopic doping percentage, we can obtain the value of Poisson ratio
from $\mu=|\varepsilon_{y}/\varepsilon_{x}|$. We find that the value
of Poisson ratio will deviate from 0.17, which means that it will
introduce some error if we use a constant value for Poisson ratio
under all environment. However, as can be seen from
Eq.~(\ref{eq_sigma2}), the Poisson ratio appears in the expression
as a factor $(1-\mu^{2})$, so the error is considerably small. For
example, we find that the largest value for Poisson ratio is 0.22 in
graphene sheet with $L=10$~{\AA} at 300 K without isotopic doping.
In this situation, the relative error is the largest, which is
$((1-0.22^{2})-(1-0.17^{2}))/(1-0.17^{2})=-2\%$. So throughout this
paper, we use a constant value for Poisson ratio, which will
introduce relative error for the Young's modulus less than $2\%$.

\textbf{Acknowledgements} JJW thanks Dr. Bo Xiong for helpful discussions. The work is supported by a Faculty Research Grant of R-144-000-173-101/112 of NUS, and Grant R-144-000-203-112 from Ministry of Education of Republic of Singapore, and Grant R-144-000-222-646 from NUS.

\end{document}